# QUANTITATIVE CHARACTERIZATION OF AMYLOID DEPOSITS IN MURINE MODELS OF ALZHEIMER DISEASE BY PHASE-CONTRAST X-RAY IMAGING

A. Politano[1*] L. Massimi[1*] N. Pieroni[1] F. Palermo[1] A. Sanna[1] M. Fratini[1] I. Bukreeva[1] C. Balducci[2*] A. Cedola[1*]

[1] TomaLab, Institute of Nanotechnology, CNR, Rome, Italy, 2 Dipartimento di Fisica, Università della Calabria, Rende, Italy,
[2] Istituto di Ricerche Farmacologiche Mario Negri IRCCS, Milan, Italy

[*] Equal contribution

Alzheimer's is a neurodegenerative disease that is the most common form of dementia, but there is still no definitive cure for this disease. The noninvasive X-ray Phase Contrast Tomography (XPCT) imaging technique was used to study brain tissues in mouse models of Alzheimer's disease, AP-PS1 and APP23. The XPCT technique enabled high-resolution imaging of brain tissues, distinguishing between different brain structures, such as amyloid deposits and neuronal cells. In addition, the XPCT technique provided detailed information on the distribution and morphology of amyloid deposits in AP-PS1 and APP23 mice putting in evidence the differences between these two models. This work demonstrates the effectiveness of this technique in supporting Alzheimer's studies and evaluating new therapeutic strategies.

**Introduction:**

Alzheimer's disease (AD) is a neurodegenerative condition characterized by progressive cognitive function and memory loss. [1-2] It is considered the most common form of dementia by the World Health Organization, affecting approximately 50 million people worldwide. While a definitive cure for AD is not currently available, various therapeutic approaches have been developed to slow down its progression.
To gain a better understanding of the disease's causes and develop novel treatments, researchers often employ animal models of Alzheimer's disease. These models provide a controlled and reproducible environment for studying the biological processes underlying the disease's progression.
Two widely used animal models for Alzheimer's disease research are the APP/PS1 and APP23 models. The APP/PS1 model involves introducing two genetic mutations, APP (Amyloid Precursor Protein) and PS1 (Presenilin 1), which are commonly associated with AD. This model exhibits rapid amyloid deposition in the brain, a key characteristic of the disease. As a result, it serves as an important model for investigating the pathological mechanisms involved in AD.
On the other hand, the APP23 model is created by introducing the APP mutation into transgenic mice. This model displays slow and progressive amyloid deposition, resembling the pattern observed in human AD. Therefore, it is utilized for studying long-term disease progression.
Both models have proven valuable in studying Alzheimer's disease, enabling researchers to explore the mechanisms underlying the disease's development and test potential therapeutic interventions. This research aims to quantitatively analyze the APP/PS1 and APP23 models, providing an overview of their main features and future prospects.
However, in order to extract useful information from animal models, it is crucial to thoroughly characterize them. The characterization process for animal models of Alzheimer's disease involves employing a wide range of techniques to study their molecular, cellular, behavioral, and functional aspects. These techniques include assessing parameters such as amyloid deposition, neurodegeneration, inflammation, synaptic dysfunction, cognitive impairment, and genetic variations.

Over the years, techniques for characterizing animal models of Alzheimer's disease have become increasingly sophisticated due to advancements in technology and scientific knowledge. For example, imaging techniques such as fluorescence microscopy, confocal microscopy, and positron emission tomography (PET) enable in vivo visualization and quantification of amyloid deposition in the brain. Additionally, DNA sequencing techniques facilitate the analysis of genetic variants in animal models. One of the most commonly employed techniques for characterizing animal models of Alzheimer's disease is the analysis of amyloid deposition. This can be achieved through imaging techniques like fluorescence microscopy, confocal microscopy, and PET, which allow for the visualization and quantification of amyloid deposition in vivo. Furthermore, the analysis of proteins such as beta-amyloid and tau can be performed using techniques like immunoprecipitation and immunofluorescence.

The analysis of neurodegeneration is another crucial technique in characterizing animal models of Alzheimer's disease. This involves examining markers of neuronal degeneration, such as microtubule-associated protein 2 (MAP2) and synapsin, which can be visualized using immunofluorescence techniques.

Inflammation, which plays a significant role in the pathogenesis of Alzheimer's disease, can be analyzed by examining pro-inflammatory cytokines like interleukin-1 beta (IL-1β) and tumor necrosis factor alpha (TNF-α). Techniques such as ELISA and real-time PCR can be utilized to measure these cytokines.

Synaptic dysfunction and cognitive impairment can be assessed through behavioral tests such as the Morris maze, nest test, and conditional fear test. These tests allow researchers to evaluate learning and memory in animal models of Alzheimer's disease.

To summarize, the techniques employed in characterizing animal models of Alzheimer's disease encompass molecular, cellular, behavioral, and functional methods. By utilizing these techniques, researchers can gain detailed insights into the animal models, thereby enhancing their understanding of the disease and facilitating the development of effective therapies. However, most of these techniques are destructive and provide 2D or 3D information at a low spatial resolution. Conventional 3D techniques like MRI and CT necessitate the use of external contrast agents to improve visibility. In neurodegenerative diseases, the use of contrast agents can alter tissue chemistry and compromise high-resolution structural analysis.

X-ray phase contrast tomography (XPCT) offers a noninvasive imaging technique that generates high-resolution 3D images of biological tissues using X-rays, without the need for external contrast agents or aggressive sample preparations.[3,4,5]

This technique relies on measuring changes in the phase and attenuation of X-rays as they pass through the sample. Recently, XPCT has been applied to the study of brain tissue for diagnosing and characterizing neurodegenerative diseases such as AD [6,7].

The APP23 mouse model, which exhibits overexpression of APP (Amyloid Precursor Protein) and develops amyloid deposits in the brain akin to those observed in AD patients, was used in this study. XPCT was employed to analyze brain tissues from both APP/PS1 and APP23 mice, aiming to evaluate the effectiveness of this technique in identifying and characterizing amyloid deposits.

**Materials and Methods:**

We used six APP23 mice and six control mice of similar age and sex. The mice were sacrificed and their brains were fixed with formalin and then subjected to XPCT imaging at the ESRF synchrotron in Grenoble. XPCT images were acquired at a resolution of 1.2 μm and were processed using the proprietary SYRMEP Tomo Project software. Phase and attenuation maps of brain tissue were calculated and regions with amyloid deposits were identified.

**Results:**

The objective of this work is to provide a detailed and accurate analysis of the number, location, and morphology of mature amyloid plaques identified by phase-contrast X-ray tomography in brains belonging to two distinct transgenic mouse models:APP23 and APP/PS1.The ultimate goal of the analysis, is to highlight the differences between the two models under investigation. It is now well established [ ] that the cerebral cortex and hippocampus are the areas most severely affected by the accumulation of amyloid protein, and furthermore, the functions in which these areas are involved correspond with those that are invalidated by the disease. It therefore seems reasonable to focus the analysis entirely in the cortical plate brain area: this, is a structure present only in embryos, and from which, during development, most of the structures that make up the cerebral cortex of an adult mouse originated. Therefore, with the term "cortical plate," we refer, in relation to adult mice, to all those structures belonging to the cerebral cortex that developed from what in the embryonic stage was the cortical plate [ ].In Figure1.a, a picture of a mouse brain is shown. Continuing still on the upper panel (a), the sagittal section of the brain is shown with the cortical plate (highlighted in yellow) and the subdivision of it into its three main structures, respectively: the isocortex (in red), the hippocampal formation (in blue) and the olfactory area (in green). In panel (b), on the other hand, the coronal, transverse and sagittal sections of the brain are represented, respectively. Highlighted are all the regions on which the analysis was conducted. From the coronal and transverse sections, it is possible to note the symmetry between the right and left hemispheres. Finally, diagram (c) illustrates the structures mentioned one by one, stating what the relationships between them are: the somatosensory and somatomotor areas constitute a portion of the isocortex; the latter, together with the hippocampal formation and the olfactory area goes instead to constitute the cortical plate. Table1 summarizes the name and abbreviation for the structures on which the plate analysis focused.

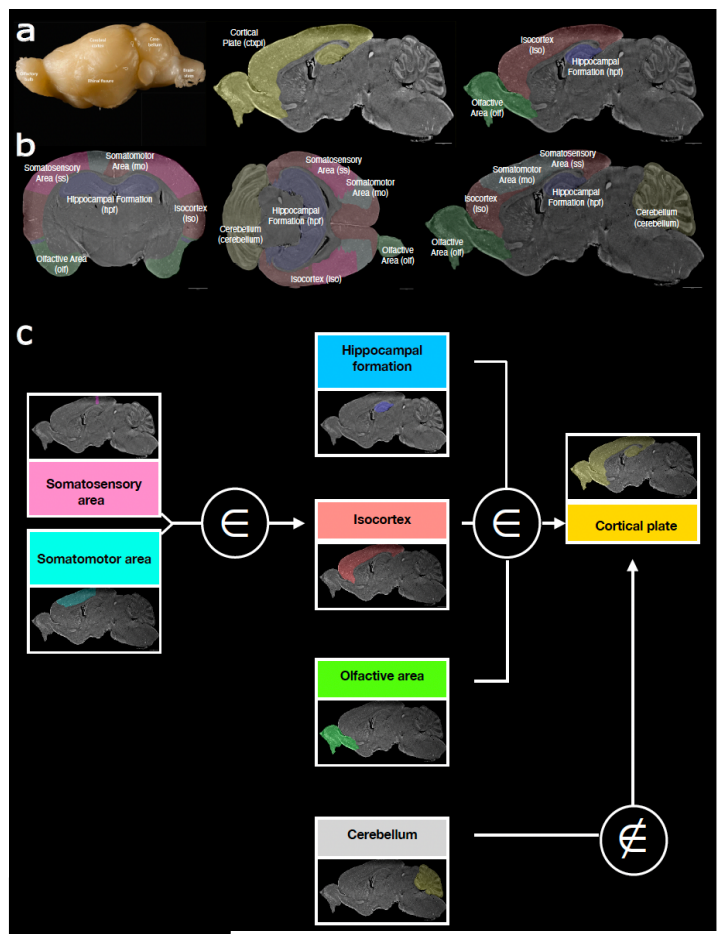

Figure 1: (a) The first image shows a photograph of a mouse brain with some of the macroareas that constitute it marked; the second presents a sagittal section of the mouse brain: the cortical plate is in evidence; in the third figure in panel (a),

on the same section, it can be seen how the cortical plate can be divided into three macro structures: the isocortex (in red), the hippocampal formation (in blue) and the olfactory area (in green). Scalebar=$1000\mu m$.(b) A coronal, transverse and sagittal section of the brain can be seen in order. Inevidence on all sections, the structures on which the analysis of plaquesamyloids was focused. Scalebar=$1000\mu m$.(c) A logical outline of the structures' belonging to the macroareas: the somatosensory and somatomotor areas are contained within the isocortex; the isocortex, hippocampal formation, and olfactory area go to make up the cortical plate. The cerebellum is invecel the only structure outside the cortical plate that has been considered.

In order to conduct an analysis of plaque distribution, it is essential to reconstruct the entire volume of the brain. This necessitates the use of 3D imaging technologies such as tomography. Prior to performing a quantitative analysis, it is possible to provide a general overview of the distribution and morphology of the plaques. As depicted in Figure 2, notable differences can already be observed between the two mouse models being investigated.

One significant difference is that in the APP/PS1 samples, a certain number of plaques can be identified in the cerebellum. However, in the APP23 model samples, no plaques are observed in that particular area.

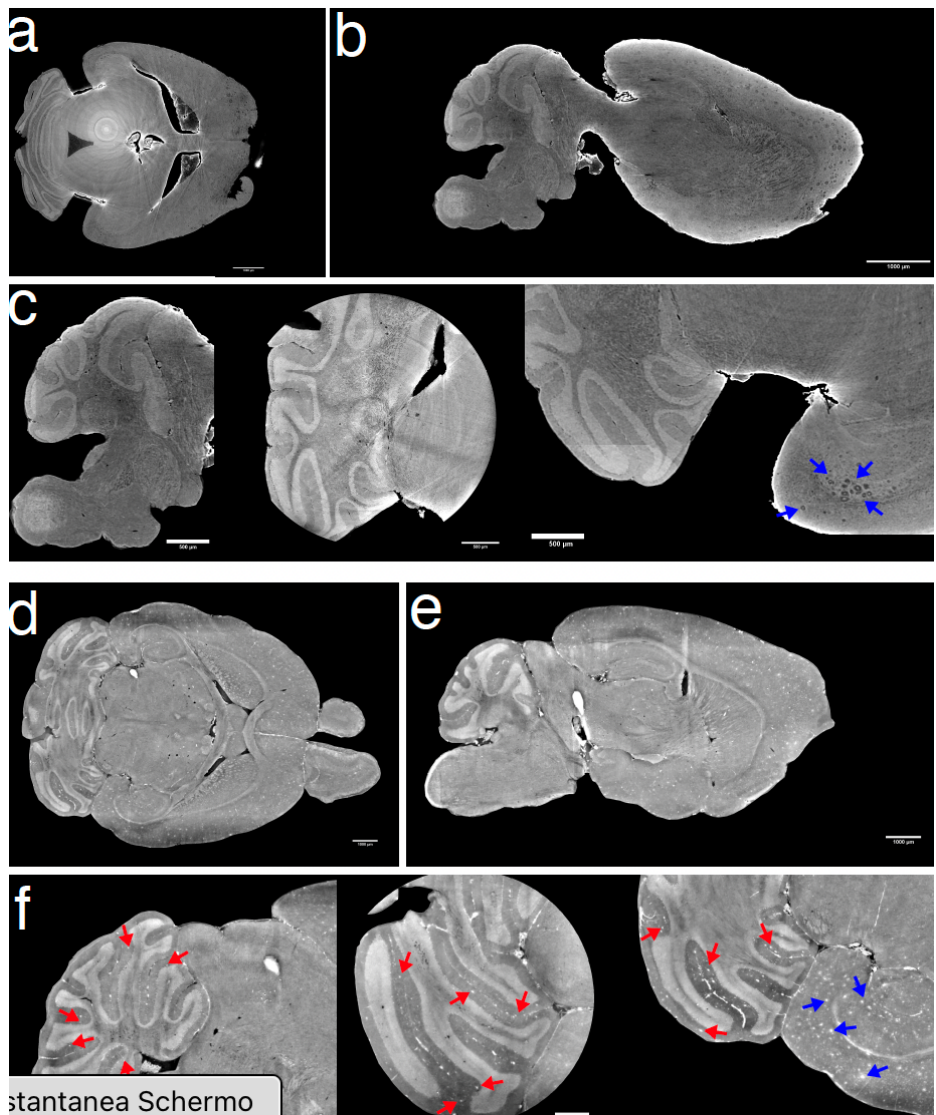

Figure 2: Amyloid plaques are marked with red arrows in case they are identified in the cerebellum,and with blue arrows otherwise.(a) Transverse section of an APP23 sample clearly shows the absence of amyloid plaques in the cerebellum. (b)Sagittal section of a sample of APP23.(c)Some insightsfrom the sections shown in (a) and (b) show well the absence of plaques in the cerebellum of APP23 samples.(d)A transverse section image of a sample of APP/PS1. (e)The sample of

APP/PS1 seen in sagittal section. (f)The insights from the projections shown in panels (d) and (e): this time the plaques are also found in the cerebellum (red arrows) as well as in the rest of the brain (blue arrows)

Providing a more detailed description, in panel (a), a transverse section of an APP23 sample is depicted, revealing the absence of amyloid plaques in the cerebellum region. Panel (b) offers a sagittal section view, confirming the same conclusion. Moving forward, panel (c) presents an enlarged view of the cerebellum region in sagittal section, followed by a region of interest (ROI) in transverse section with a pixel size of 0.7μm. Additionally, an enlargement of a transverse section illustrates a portion of the cerebellum and hippocampal formation, with blue arrows indicating the presence of amyloid plaques in the hippocampus. Across all panels, it becomes evident that there is a lack of plaques in the cerebellum, while a substantial number are present in the cortical plate region.

On the other hand, panels (d), (e), and (f) provide corresponding images for the APP/PS1 model. In these images, amyloid plaques in the cerebellum are clearly visible in both transverse (panels d and f) and sagittal (panel e) sections. Compared to the surrounding areas, the cerebellum is easily recognizable, making this difference between the two models highly apparent. Based on this initial qualitative evidence, it was decided not to perform cerebellum segmentation for the APP23 model. Conversely, in the APP/PS1 samples, the presence of a large number of plaques in this area was immediately noticeable, leading to its inclusion in the analyzed structures.

Continuing with the assessment of plaque distribution, from a qualitative standpoint, it appears that plaques in the APP23 model tend to cluster more closely together compared to the more uniform distribution observed in the APP/PS1 model. This distinction is noticeable in Figure 2, where the transverse (panel a) and sagittal (panel b) sections of an APP23 sample are compared with the corresponding sections of an APP/PS1 sample (panels d and e). Specifically, when comparing the transverse sections, it can be observed that plaques in the APP23 model are larger and concentrated in close proximity, creating zones of higher plaque density interspersed with areas of lower or almost no plaque density. In contrast, the plaques in the APP/PS1 model appear smaller and exhibit a more consistent distribution along the entire extent of the cortical plate. This distribution difference can be further appreciated in the images shown in Figure 3, where coronal and transverse sections are presented for both models. Clusters of plaques are circled in red, while small accumulations or isolated plaque findings are indicated by blue arrowheads. While the APP23 model displays areas with significant plaque clusters, the APP/PS1 model exhibits a more homogeneous distribution of amyloid presence throughout the cortical plate.

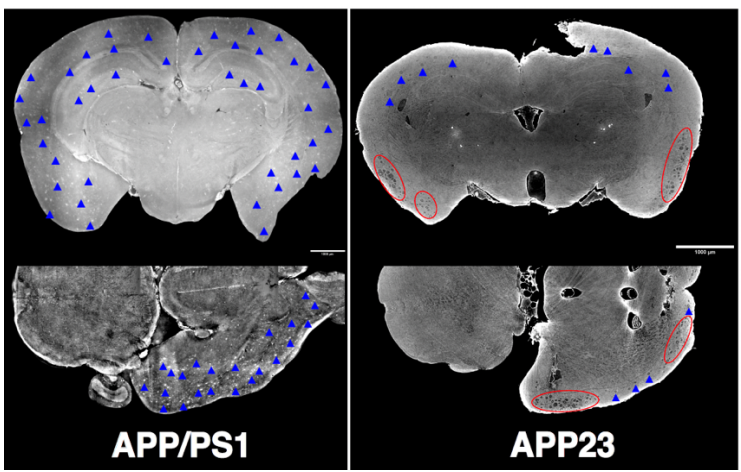

Figure 3: Clusters of plaques are circled in red, while isolated presences or small clusters are indicated by blue arrowheads. The coronal and transverse sections of the APP/PS1 model show a very homogeneous distribution, while in the case of APP23there are several clusters to report.

Upon closer examination of the images, it becomes evident that the plaques in the two models exhibit significant structural differences as well. An initial analysis of high-resolution images in Figure 4 reveals interesting disparities in the appearance of the plaques. The left images in panels (a) and (b) showcase a portion of the hippocampal formation in cross-section for both analyzed models. The region of interest (ROI) was investigated using a pixel size of 0.7μm. However, for better management of the image stack, the high-resolution slices were rescaled by a factor of 1/2 in all dimensions, resulting in voxel sizes of 1.4μm. The plaques identified in the brains of tg APP23 mice exhibit an outer surface enclosing a higher-density structure. Internally, they possess a denser structure compared to the surrounding tissue, which appears as a spherical structure with lower density, resembling an empty space or vesicle. Conversely, in the brains of tg APP/PS1 mice, the plaques appear uniformly white, indicating a high and homogeneous density, with no apparent difference between the inner and outer surfaces.

Figure 4, in panel (a), presents two transverse sections side by side for each model. The left images display some plaques located in the hippocampal formation, while the right images depict plaques found in the isocortex. Morphologically, there don't appear to be substantial differences between the deposits in different substructures of the cortical plate. This implies that plaque analysis can be conducted on any ROI focusing on a portion of the cortical plate. Additionally, panel (b) demonstrates that the appearance of plaques in the respective models remains consistent regardless of whether they are observed in transverse, sagittal, or coronal sections. This observation is further highlighted in the magnified view of a single plaque shown in panel (c). Although most plaques exhibit a spherical shape, there are also some elongated deposits, marked with red arrows in the figure. An initial qualitative analysis suggests that objects with a particular orientation are not expected geometrically. Furthermore, panel (c) includes histological images for each model alongside the tomographic images, confirming the structural difference between the plaques of the APP23 and APP/PS1 models as previously highlighted.

Quantitative Analysis:
Once all the brain structures of interest have been isolated using high-resolution images, the structure of the amyloid plaques can be further investigated. The pixel size used for image acquisition is 0.7μm. However, as mentioned during the qualitative analysis, a rescaling process was employed to better manage the image stack, resulting in an effective pixel size of 1.4μm. Similar to the brain areas, individual plaques need to be segmented to perform a quantitative analysis. The segmentation procedure follows the same manual approach as described for the 3μm images, where the shape of the plaque of interest is outlined on each individual slice. The decision to opt for manual segmentation is due to numerous elements in each slice generating a contrast similar to that of the plaques, making threshold segmentation challenging. Since the average size of amyloid plaques is much smaller than that of the brain areas, a smaller "segmentation step" needs to be chosen instead of segmenting every 60μm slice. In the images in Figure 4.8, the variation of plaques is observed every 5 slices in a region of interest within the transverse section of an APP23 model sample, specifically within the hippocampal formation. Some plaques that are significant for the analysis are labeled in red, and their development along the transverse section axis can be tracked across the succession of images. Each slice has a thickness of 1.4μm, and it is apparent that most plaques extend beyond the chosen pitch. Hence, a segmentation is performed every 5 slices along the transverse section. While the localization of plaque locations was conducted across the entire brain, their segmentation was limited to a few regions of interest situated in the hippocampal formation of the APP23 and APP/PS1 samples. There are two main reasons for this approach: firstly, the number of plaques is exceedingly high, and secondly, as demonstrated in the qualitative analysis (Figure 4.4), there don't appear to be substantial morphological differences between plaques originating from different brain structures, except for those found in the cerebellum of APP/PS1 samples, as indicated by Massimi et al. [ ]. However, these cerebellar plaques are excluded from the comparison between the two models due to their absence in the APP23 samples.

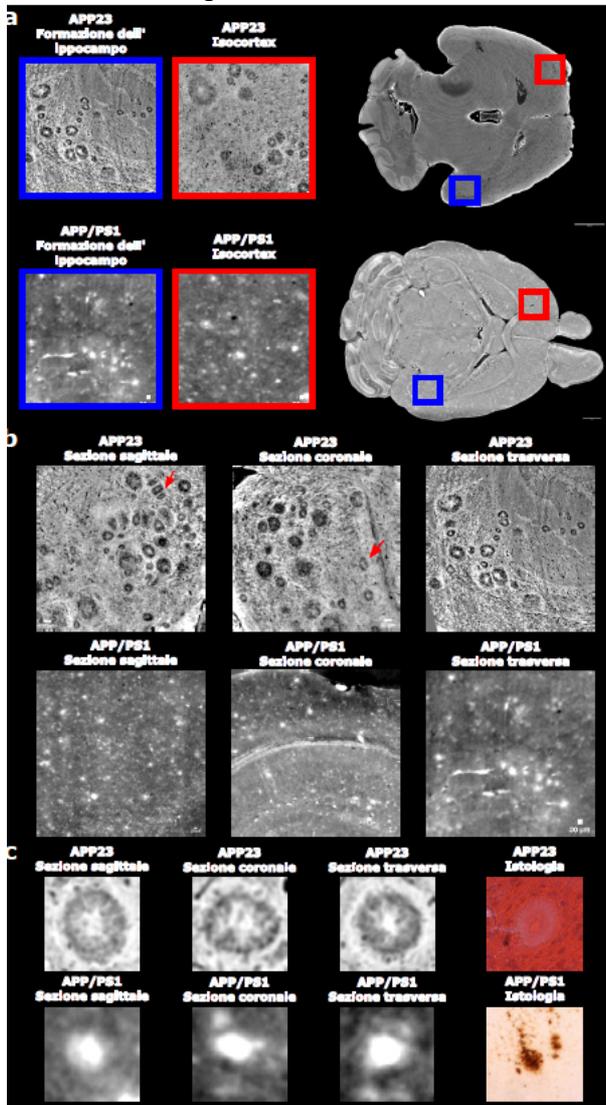

Figure 4: (a) Comparison of plaques in the hippocampal formation and isocortex of theAPP23 model and the APP/PS1 model. Pixel size= 1.4$\mu m$. Considering the deposits detected in thecortical plate, from a morphological point of view, the amyloid plaques do not show significant differences depending on the substructure in which they form. (b) In succession: theagittal, coronal and transverse sections of the amyloid plaques in the two models. Barring a few exceptions(marked with a red arrow), most plaques have a circular structurefrom whichever section one chooses to look at them. This is reflected in volumes that are fundamentallyapproximable as spherical. (c) A magnification on a single plaque of the APP23e model of APP/PS1 seen in all three sections. On the right, however, a comparison histological image depicting one plaque from each of the two models

By employing the segmentation procedures described earlier, it is now possible to progress from simple qualitative observations to more mathematically rigorous results. One such analysis involves calculating the volume (V) occupied by the cerebral structures. This can be achieved by multiplying the volume of a voxel (in this case, a cube with a side length of 3μm) by the total number of pixels comprising the virtual volume. Consequently, both the volume of the entire brain and that of the substructures can be derived. Figure 5 presents the histogram displaying the obtained results (panel c), accompanied by an illustration showcasing all the structures whose volume was calculated on a tomographic slice (panels a and b).

Specifically, the following observations are made for the APP23 and APP/PS1 models, respectively:

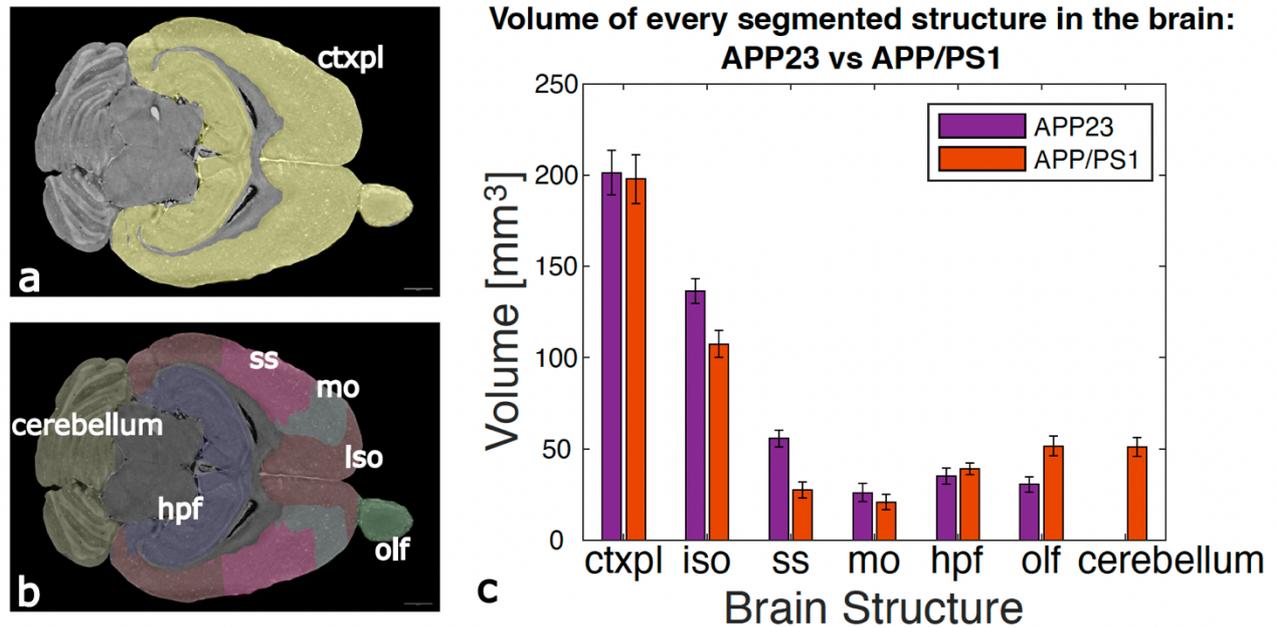

Figure 5: (a) Highlighting the cortical plate in transverse section.(b) The brain structures whose volumes are displayed in panel (c).(c) The volumes of the segmented structures. In addition to the total volume of the cortical plate (ctxpl), that of the isocortex (iso) and two of its cortical areas: the somatosensory(ss) and the somatomotor (mo) are also shown. Completing the graph are the volume of the hippocampal formation (hpf) and the olfactory area (olf). The last position is reserved for the cerebellum(cerebellum): it is a structure outside the cortical plate, however, as already demonstrated infigure 4.2, in the samples of the APP/PS1 model, numerous amyloid plaques were identified and, since the focus of the paper is precisely to analyze the distribution of these plaques, the segmentation of the cerebellum was also carried out, but only for theAPP/PS1 model: the APP23 model does not appear to be affected by the amyloid phenomenon in this area.

After segmenting the brain areas, the amyloid plaques present throughout the entire brain were identified and marked. Due to the presence of elements with plaque-like contrast that complicated threshold segmentation, manual segmentation was performed once again. As mentioned in the qualitative analysis and segmentation section, the amyloid plaques in the samples have a thickness greater than 3μm, which is the spatial resolution of the stack. This implies that each plaque is contained in more than one slice, and counting them requires checking each plaque individually to ensure it has not already been counted in the previous slice. However, this operation is time-consuming and requires considerable concentration.

To address this challenge, an analysis was conducted to determine approximately how many slices are needed to contain a plaque by examining the average diameter of a substantial pool of plaques (approximately 150 plaques for each model). Based on this analysis, it was deemed reasonable to count the plaques on one slice every 36μm (12 slices) instead of checking each slice individually. This choice aligns with the preference to analyze images representing thicker volumes, as histologic and tomographic results are often synergistically used to analyze tissue. Since histologic images are obtained from thicker tissue sections (generally greater than 30μm), they need to be compared with tomographic images referring to an equal thickness.

It should be noted that this counting method may result in the omission of some plaques smaller than the count step used (36μm) and entirely contained within the skipped section. However, the focus of plaque counting is primarily on their distribution and density, while a more detailed analysis of plaque size, shape, and morphology is reserved for a higher resolution investigation in the subsequent section. Despite the possibility of missing some plaques, the large number of plaques included in the analysis still allows for reliable statistics regarding their distribution.

Counting the plaques was performed on the projections (either max or min) along 12 slices, as mentioned earlier, starting from the transverse section of the stack. The result yielded the position (x, y) of the plaques in the projection plane and a position along the z-axis. By utilizing the previously created masks, it became possible to determine the number of plaques belonging to each specific structure. However, since the masks were generated in coronal section and the plaque localization was conducted in transverse section, a new viewing section of the stack needed to be reconstructed. This reslice operation allowed the stack to be derived in transverse section from the coronal section, as shown in Figure 6.

The plates were counted using projections based on intensity values. To superimpose the plaque regions of interest (ROIs) on the mask, a projection was made every 36μm (12 slices) of the mask (either maxima or minima, as it is indifferent since binary objects are considered). The process of creating the corresponding transverse mask through reslicing was repeated for all structures (cortical plate, isocortex, somatosensory area, hippocampal formation, olfactory area) and all samples (two samples for APP23 and two for APP/PS1). Once the masks were prepared, the plaque coordinates were overlaid on them. Figure 7 provides a schematic representation of this procedure using the example of the cortical plate.

Finally, several relevant quantities of the amyloid plaques were calculated. For the purposes of the present application, the key parameters of interest include the coordinates of the plaque and the pixel value at the identified plaque center. Based on this value, it is possible to determine whether the plaque is located on a specific structure or not.

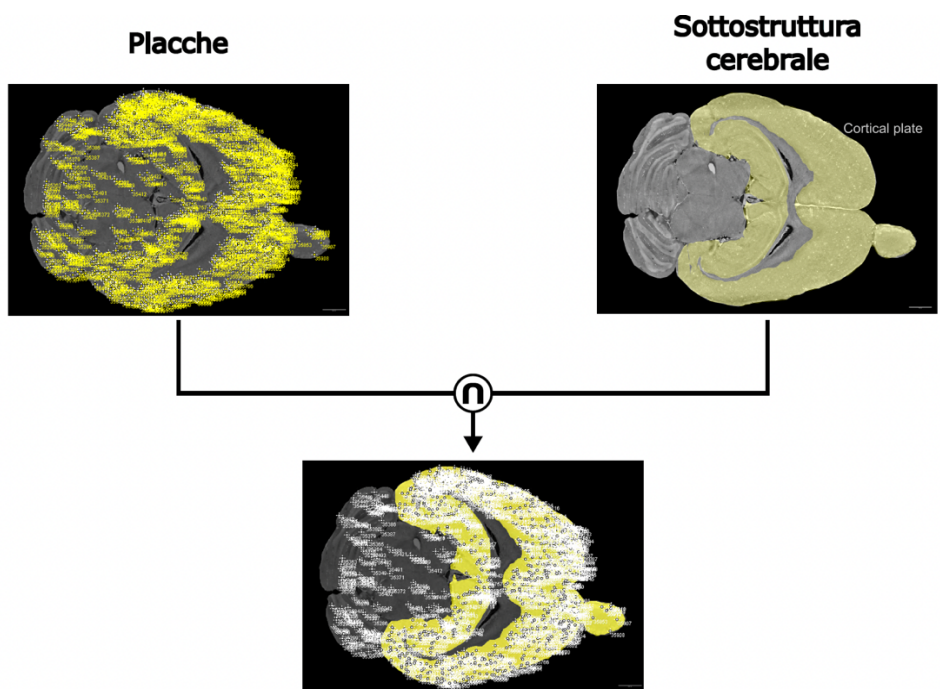

Figure 6: Plaques were counted on the volume projections: a file was created containing all the selections corresponding to the identified amyloid plaques. Once the brain substructure is selected on which one wants to know how many plaques were found, an intersection is made between mask and file with the coordinates of the plaques. The operation is schematized here using the cortical plate example, but it is repeated on each segmented area cerebral.

Table 2 summarizes the results obtained from the plaque counting analysis. It confirms that the number of plaques is higher in the APP/PS1 model compared to the APP23 model, which is consistent with prior knowledge of these models. The APP/PS1 model is designed to develop plaques similar to human plaques at an accelerated rate, resulting in a larger number of plaques within a shorter timescale. On the other hand, the APP23 model exhibits a lower plaque count. These findings align with existing literature on these models [11, 24, 20].

The distribution of plaques within the brain structures was also examined. The majority of amyloid plaques in both models (more than 80%) were found in the cortical plate. In the APP23 model, 95% of the plaques belonged to the cortical plate, with the remaining plaques distributed in fibrous tracts or other structures. In the APP/PS1 model, 84% of the plaques were located in the cortical plate, while the remaining plaques were evenly divided between the cerebellum and other structures. These observations validate previous knowledge of these models and are consistent with the literature [11, 24, 20].

Further analysis focused on the cortical plate substructures, specifically the isocortex, hippocampal formation, and olfactory area. The results showed that most of the plaques were found in the isocortex, while the hippocampal formation and olfactory area shared the remaining plaques. This finding aligns with the initial hypothesis that a significant number of plaques would be present in the cerebral cortex. Additionally, the APP23 model exhibited a lower percentage of plaques in the hippocampus compared to the APP/PS1 model.

The involvement of somatosensory and somatomotor areas in Alzheimer's disease (AD) was also investigated. While there is debate regarding the role of these areas in AD, previous studies have suggested their involvement in the advanced stages of the disease [25, 26, 27, 28]. Abnormalities in neuronal responses to stimuli have been observed in the somatosensory area [29], indicating its potential early involvement in the progression of AD. These findings further support the hypothesis that somatosensory cortex may be affected earlier than expected in AD, with behavioral and functional consequences.

In the analysis of plaque distribution within a certain radius around each plaque, the average number of plaques was found to be higher in the APP23 model samples compared to the APP/PS1 model samples. This observation aligns with the qualitative analysis, which revealed that plaques in the APP23 model tend to cluster and agglomerate, resulting in fewer but densely occupied brain areas. In contrast, APP/PS1 plaques exhibited a more homogeneous distribution. The growth of plaque numbers linearly correlated with volume, and as the search radius increased, the number of plaques in the APP23 model reached a plateau earlier than in the APP/PS1 model. The total number of plaques was significantly higher in the APP/PS1 model, leading to a slower change in concavity compared to the initial spread observed in the APP23 model.

Overall, these findings provide further insights into the distribution and characteristics of amyloid plaques in the examined brain structures in the APP23 and APP/PS1 mouse models, supporting previous knowledge and literature on these models.

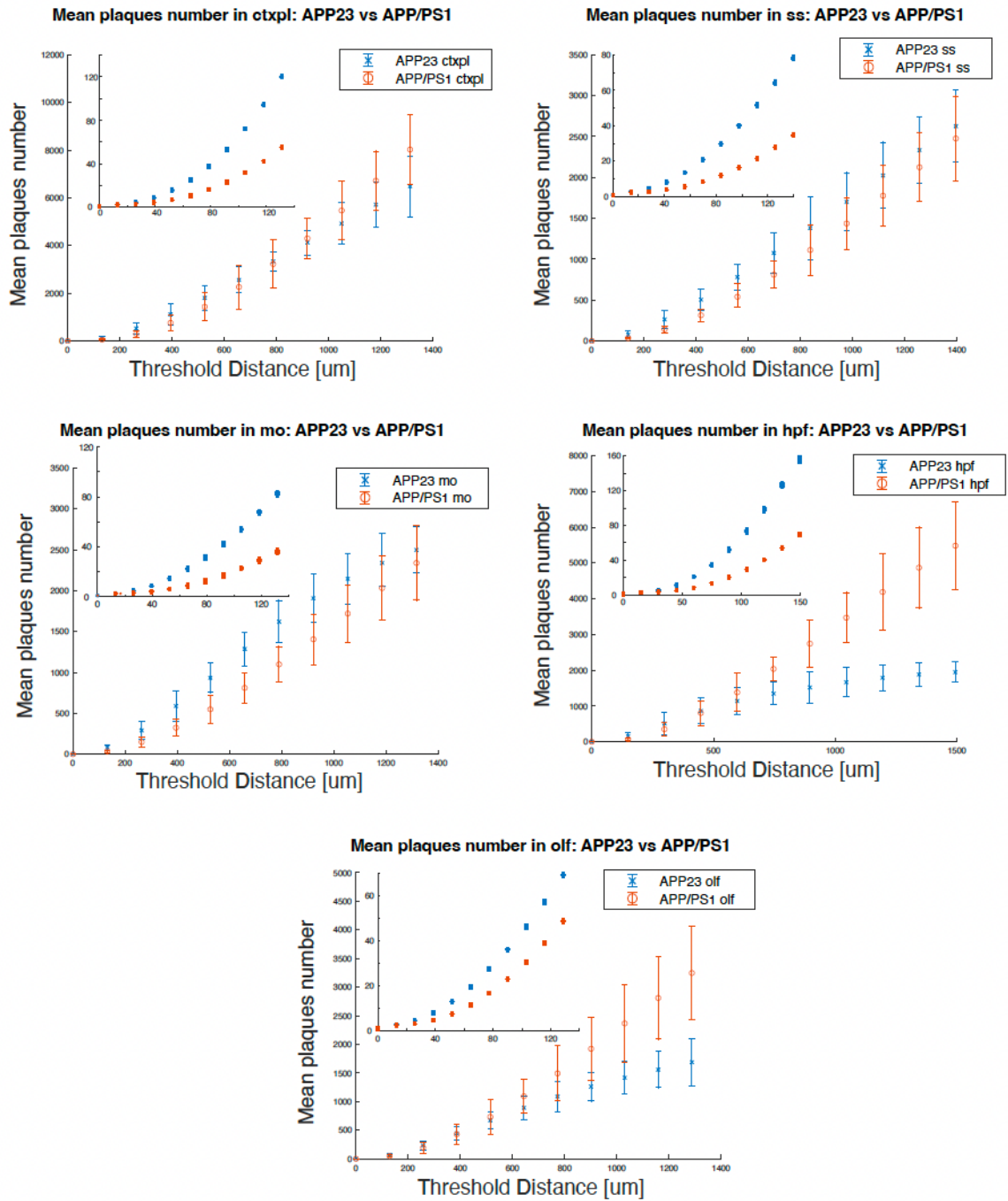

Figure 7: The trend of the average number of plaques as the threshold radius changes. In the insets, the distribution of the number of plaques for distance values within 10times the value of the average minimum distance is presented. It can be seen that all structures initially record a higher number for the APP23 model, as the plaques in these samples are more clustered. As the size of the sphere within which to count the plaques grows, however, it becomes more and more evident that the number of plaques is greater in the APP/PS1 model, consequently the trend for APP23 changes laconcavity and tends toward the limiting value, which coincides with the total number of plaques in that structure, while that for APP/PS1 continues to grow, as the number of plaques grows initially less rapidly but more steadily.

The shape analysis of plaques in the APP23 and APP/PS1 models revealed some differences in their characteristics. In the APP/PS1 model, the plaques appeared predominantly spherical but with slightly more irregular shapes. Some plaques exhibited jagged edges and a more varied structure, including a significant percentage of slightly oblong objects. On the other hand, the plaques in the APP23 model showed a higher degree of regularity and spherical shape.

Figure 8 illustrates the segmentation of two specific plaques, one from the APP23 model and the other from the pool of segmented plaques in the APP/PS1 model. The plaque from the APP23 model had a roundness of 0.76 and a sphericity of 0.86, indicating a relatively regular and spherical three-dimensional structure. In contrast, the plaque from the APP/PS1 model exhibited a roundness of 0.70 and a lower sphericity of 0.61, suggesting an elongated and angular shape.

These findings align with the initial qualitative analysis, which noted the differences in contrast and surface characteristics between the plaques of the two models. However, the exact reason for the different contrast on the outer surface of the plaques in the two models remains an open question that requires further investigation.

Overall, the shape analysis provides additional insights into the structural characteristics of plaques in the APP23 and APP/PS1 models, highlighting their differences in shape, regularity, and surface characteristics.

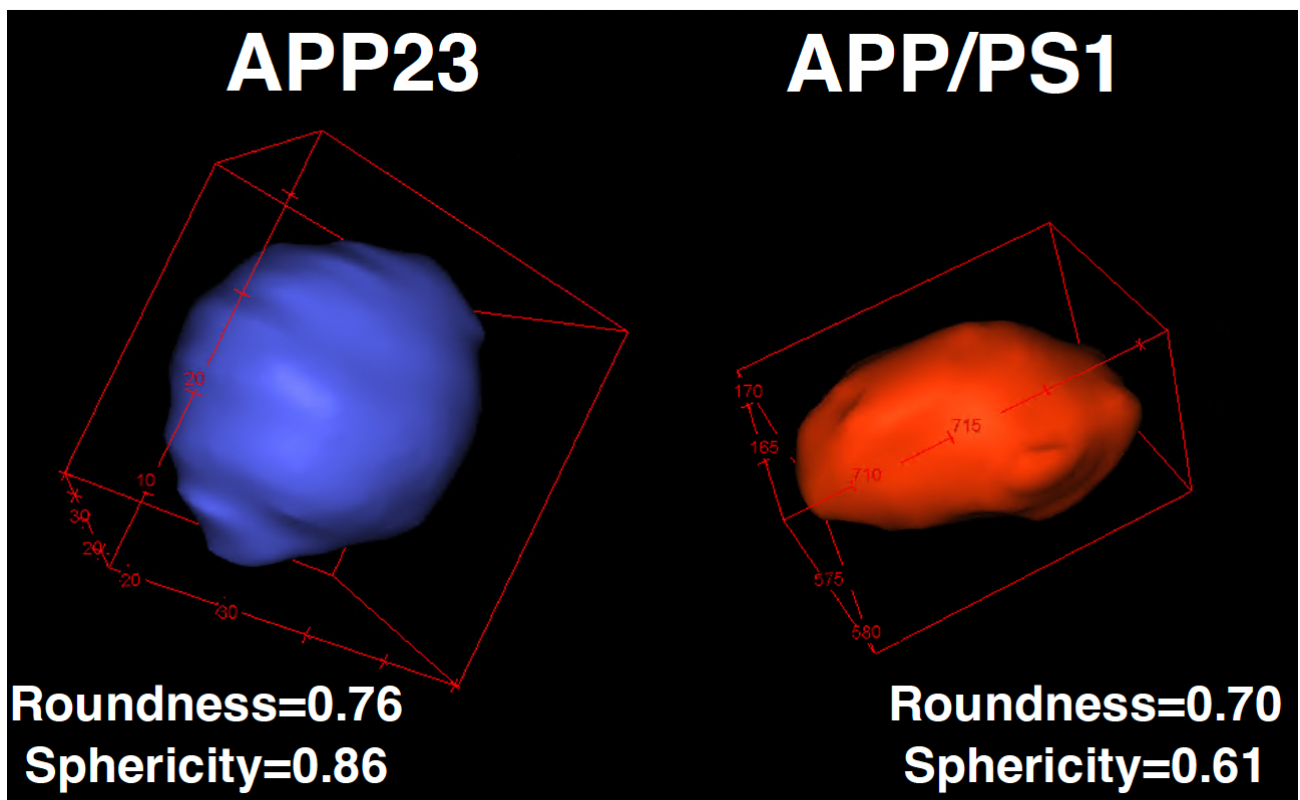

Figure 8: 3D image of a plaque belonging to the APP23 model and one belonging to the APP/PS1 model. The examples shown, were chosen to highlight someemorphological differences found between the plaques of the two different transgenic lines

In conclusion, the phase-contrast X-ray tomography imaging method proved to be highly effective for acquiring measurements of amyloid plaques in the APP23 and APP/PS1 models. The technique allowed for the analysis of the entire brain volume without compromising its structure, providing high-resolution quantitative data that would be challenging to obtain through other investigative methods.

The qualitative analysis revealed noticeable differences between the plaques in the two models. The plaques in the APP23 model appeared larger in size compared to those in the APP/PS1 model. Additionally, the contrast generated by the plaques differed, with the plaques in the APP23 model exhibiting a significantly denser inner structure than the outer structure, while the plaques in the APP/PS1 model showed no difference between the inner and outer regions. These observations were further supported by histology, which confirmed the presence of amyloid plaques in the cerebellum of the APP/PS1 model but not in the APP23 model.

The distribution of plaques was another significant difference between the two models. The plaques in the APP23 model exhibited a greater tendency to cluster, whereas the distribution of plaques in the APP/PS1 model was more homogeneous and consistent throughout the brain structures.

The morphological analysis highlighted additional differences. The APP/PS1 model showed an elog-normal distribution of plaque volume and surface area, consistent with previous studies. In the APP23 model, the plaques exhibited two more likely values of volume. Overall, both models demonstrated a proportional relationship between the growth rate of plaque surface area and volume.

Regarding plaque shape, both models predominantly exhibited spherical plaques. However, the plaques in the APP/PS1 model displayed a broader distribution, indicating the presence of plaques with more elongated or flatter shapes. The roundness analysis indicated that the plaques in the APP23 model were slightly rounder compared to those in the APP/PS1 model, supporting the initial observations made during the qualitative analysis.

These findings contribute to a better understanding of the characteristics and distribution of amyloid plaques in the APP23 and APP/PS1 models of Alzheimer's disease. The quantitative measurements obtained through phase-contrast X-ray tomography provide valuable insights into the structural properties of plaques and highlight the differences between the two models, furthering our knowledge of the disease's progression.